\documentclass[prl,showpacs,showkeys]{revtex4}

\usepackage{graphicx}
\usepackage{amsmath}
\usepackage{bm}

\begin{document}

\title{Optimal Dynamical Range of Excitable Networks at Criticality}
\author{Osame Kinouchi}
\email{osame@ffclrp.usp.br}
\thanks{Corresponding author}
\affiliation{Departamento de F{\'\i}sica e Matem\'atica, Faculdade de
  Filosofia, Ci\^encias e Letras de Ribeir\~ao Preto, Universidade de S\~ao
  Paulo, Av. dos Bandeirantes 3900, 14040-901, Ribeir\~ao Preto, SP, Brazil}
\author{Mauro Copelli}
\email{mcopelli@df.ufpe.br}
\affiliation{Laborat\'orio de F\'{\i}sica Te\'orica e Computacional,
  Departamento de F\'{\i}sica, Universidade Federal de Pernambuco,
  50670-901 Recife, PE, Brazil}

\begin{abstract}
A recurrent idea in the study of complex systems is that optimal
information processing is to be found near bifurcation points or phase
transitions. However, this heuristic hypothesis has few (if any)
concrete realizations where a standard and biologically relevant
quantity is optimized at criticality.  Here we give a clear example of
such a phenomenon: a network of excitable elements has its sensitivity
and dynamic range maximized at the critical point of a non-equilibrium
phase transition. Our results are compatible with the essential role
of gap junctions in olfactory glomeruli and retinal ganglionar cell
output. Synchronization and global oscillations also appear in the
network dynamics. We propose that the main functional role of
electrical coupling is to provide an enhancement of dynamic range,
therefore allowing the coding of information spanning several orders
of magnitude. The mechanism could provide a microscopic neural basis
for psychophysical laws.
\end{abstract}

\pacs{87.18.Sn, 87.19.La, 87.10.+e, 05.45.-a, 05.40.-a}
\keywords{Gap junction, Ephaptic interaction, Olfaction, Retina,
Excitable media, Neural code, Dynamic range, Cellular automata}

\maketitle

Psychophysics is probably the first experimental area in
neuropsychology, having been founded by physiologists (Helmholtz and
Weber) and physicists (Fechner, Plateau, Maxwell and Mach) in the
middle of the nineteenth century~\cite{Stevens}. Its aim is to study
how physical stimuli transduce into psychological sensation, probably
the most basic mind-brain problem. How to relate such high-level
psychological phenomena to low-level neurophysiology is a task for the
twenty-first century brain sciences.  

As the intensity of physical stimuli (light, sound, pressure, odorant
concentration etc) varies by several orders of magnitude,
psychophysical laws must have a large dynamical range. Although this
has been extensively verified experimentally at the
psychological~\cite{Stevens} and
neural~\cite{Wachowiak01,Angioy03,Fried02} level, little work has been
done regarding the mechanism that produces such psychophysical
laws~\cite{Cleland99,Copelli02,Copelli05b,Copelli05a}. However, two
main ideas seem to be consensual: (1) nonlinear transduction must be
done at the sensory periphery to prevent early
saturation~\cite{Stevens}; and (2) the broad dynamic range is often a
collective phenomenon~\cite{Stevens,Cleland99,Copelli02}, because
single cells usually respond in a linear saturating way with small
ranges~\cite{Reiser01,Tomaru05}.

Two nonlinear transfer functions have been widely used to fit
experimental data, both in psychophysics and in neural response: the
logarithm function $F(S) = C \log S$ (Weber-Fechner law) and the power
law function $F(S) = C S^m$ (Stevens Law), where $S$ is the stimulus
level, $C$ is a constant and $m$ is the Stevens exponent. Later, other
functions have been proposed to fit data with more extended input
range, and to account for sensory saturation, in particular the Hill
function $F(S) = F_{max}S^m/(S^m+S_0^m)$ where $F_{max}$ is the
saturation response and $S_0$ is the input level for half-maximum
response. Notice that, because both Hill and Stevens functions have a
power law regime, it is natural to denote the exponents by the same
parameter $m$.

Some authors have tried to derive such phenomenological laws from the
structure of natural signals~\cite{Chater99}. This type of work may
furnish an evolutionary motivation for biological organisms to
implement or approximate such laws. However, there is no consensual
view on this theoretical aspect, and it does not provide a neural
basis for implementing the psychophysical laws. 

In contrast, our statistical physics approach to neural
psychophysics~\cite{Copelli02,Copelli05b,Copelli05a,Furtado06} shows
how Hill-like transfer functions may arise in biological excitable
systems: they are not put into the models by hand, nor mathematically
derived from {\it a priori\/} reasoning, but appear as a cooperative
effect in a network of excitable elements. At differente biological
levels, these elements may be interpreted as whole neurons, excitable
dendrites, axons, or other subcellular excitable units.  We show that
a network of excitable elements, each with small dynamic range,
presents a collective response with broad dynamic range and high
sensitivity. Even more interesting, we find that the dynamic range is
maximized if the spontaneous activity of the network corresponds to a
critical process. This is compatible with recent findings of critical
avalanches in {\it in vitro\/} neural
networks~\cite{Beggs03,Haldeman05} and provides a clearcut example of
optimal information processing at criticality~\cite{Langton,Chialvo04,
bak}. The model also has other dynamical features and permits us to
make a testable prediction.

In previous work we have introduced the idea that excitable waves in
active media provide a mechanism for strong nonlinear amplification
with large dynamic range.  This has been shown in simulations with
one-dimensional~\cite{Copelli02} and two-dimensional~\cite{Copelli05b}
deterministic cellular automaton models, as well as with
one-dimensional networks of coupled maps and Hodgkin-Huxley
elements~\cite{Copelli05a}. Analytical results have recently been
obtained for the one-dimensional cellular automaton model under the
two-site mean-field approximation~\cite{Furtado06}. In this paper, we
study a very different system where the activity propagation is
stochastic and the electrical synapses form a random network. This
latter case seems to be a more realistic topology for, say, the
olfactory intraglomerular network of excitable dendrites coupled by
gap junctions~\cite{Kosaka05b,Migliore05,Christie05}. We find a whole
new phenomenology related to a non-equilibrium phase transition to
re-entrant activity present in this model.

In the present model, each excitable element $i = 1,\ldots, N$ has $n$
states: $s_i = 0$ is the resting state, $s_i = 1$ corresponds to
excitation and the remaining $s_i = 2,\ldots, n-1$ are refractory
states. There are two ways for the $i$th element to go from state $s_i
= 0$ to $1$: (1) owing to an external stimulus, modelled here by a
Poisson process with rate $r$ (which implies a transition with
probability $\lambda=1-\exp(-r\Delta t)$ per time step); (2) with
probability $p_{ij}$, owing to a neighbour $j$ being in the excited
state in the previous time step. Time is discrete (we assume $\Delta t
= 1$ ms) and the dynamics, after excitation, is deterministic: if $s_i
= 1$, then in the next time step its state changes to $s_i =2$ and so
on until the state $s_i = n-1$ leads to the $s_i = 0$ resting state,
so the element is a cyclic cellular automaton~\cite{Marro99}. The
Poisson rate $r$ will be assumed to be proportional to the stimulus
level $S$ (for example, the odorant concentration in olfactory
processing). Notice that each element receives external signals
independently, that is, we have a Poisson process for each element
(modelling the arrival of axonal inputs from different receptor
neurons). 

\begin{figure}[!h]
\includegraphics[width=\columnwidth]{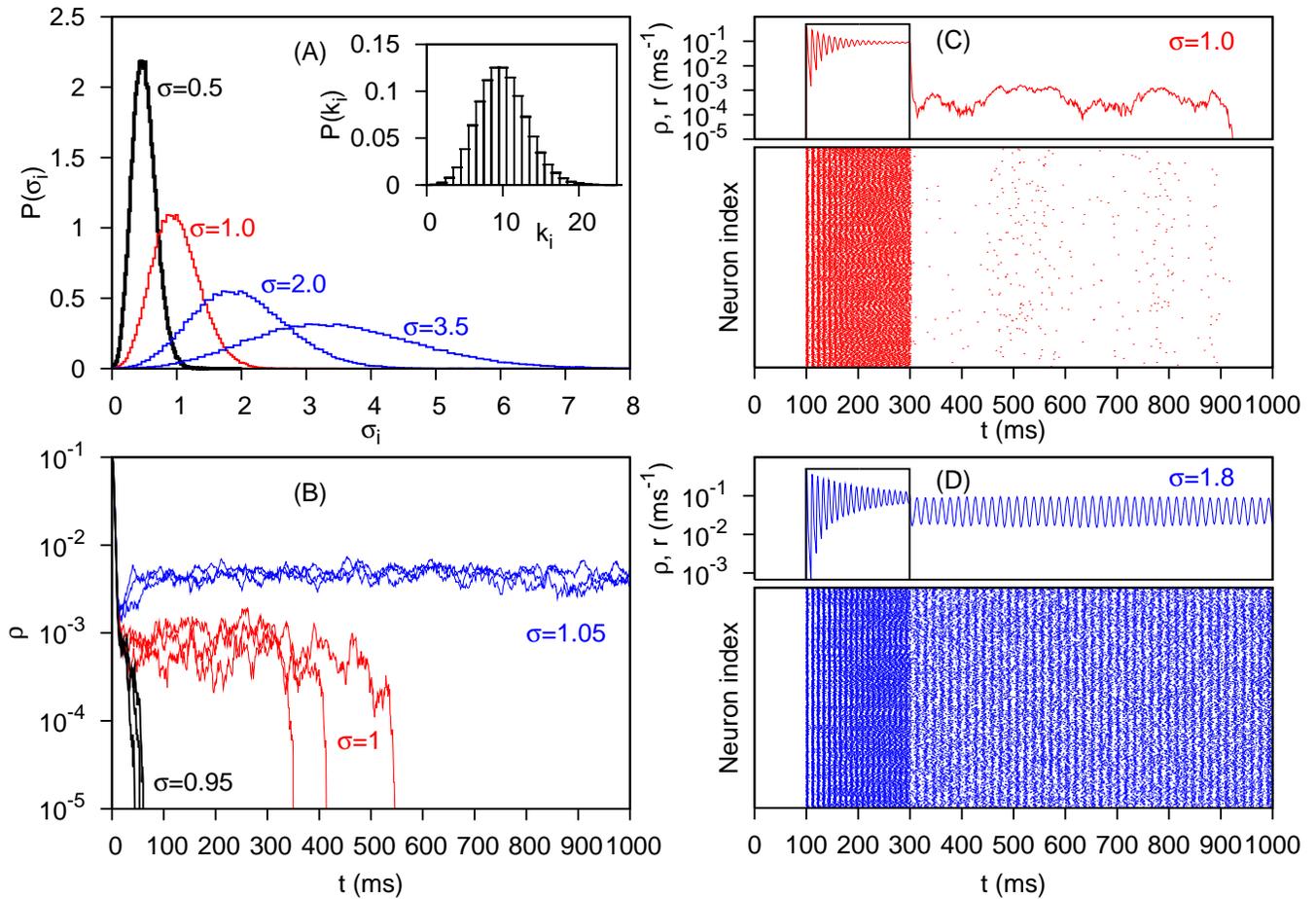}
\caption{Network characterization and density of active
sites. Simulations with $N = 10^5$ sites, $K = 10$ and $n = 10$
states. (A) Probability density function of local branching ratio and
(inset) connectivity (``degree'') distribution; (B) Instantaneous
density of active sites for subcritical (black), critical (red) and
supercritical (blue) branching parameters as functions of time (three
different runs for each case); (C), (D) Instantaneous density (of all
sites; upper panels) and raster plot (of $10^3$ randomly chosen sites;
lower panels) in response to a square pulse of stimulus ($r =
0.5$~ms$^{-1}$ for 100~ms $\leq t \leq$ 300~ms, null otherwise) for
critical (C) and supercritical (D) branching parameters.}
\end{figure}

The network with $N$ elements is an Erd{\H o}s-R\'enyi undirected
random graph, with $NK/2$ links being assigned to randomly chosen
pairs of elements. This produces an average connectivity $K$ where
each element $i$ $(i = 1,\ldots, N)$ is randomly connected to $K_i$
neighbours. The distribution $P(K_i)$ of neighbours is a binomial
distribution with average $K$ (see Fig 1A).  The probability that an
active neighbour $j$ excites element $i$ is given by $p_{ij}$, a
random variable with uniform distribution in the interval
$[0,p_{max}]$. The weights are symmetrical ($p_{ij} = p_{ji}$) and are
kept fixed throughout each simulation (``quenched disorder''). This
kind of coupling models electric gap junctions instead of chemical
synapses because it is fast and bi-directional. However, symmetry is
not a necessary ingredient, because similar results are obtained in
asymmetrical networks. Note that we are not assuming that gap
junctions have a stochastic dynamics, but only that, because other
internal factors and noise are also present, a probabilistic account
of the excitation process may be more realistic.

The local branching ratio $\sigma_j = \sum_i^{K_j} p_{ij}$ corresponds
to the average number of excitations created in the next time step by
the $j$-th element~\cite{Haldeman05}. The distribution $P(\sigma_j)$
of local branching ratios (a bell-shaped distribution with average
$\sigma=\left<\sigma_i\right>$) is shown in Fig.~1A. The average
branching ratio $\sigma$ is the relevant control parameter. In the
simulations, we set $\sigma$ by choosing $p_{max} = 2\sigma/K$ and
keeping $\sigma< K/2$.

The network instantaneous activity is the density $\rho_t$ of active
($s = 1$) sites at a given time $t$. We also define the average
activity $F=T^{-1}\sum_{t=1}^T \rho_t$ where $T$ is a large time
window (of the order of $10^3$ time steps). Typical $\rho_t$ curves in
the absence of stimulus ($r = 0$) are shown in
Fig.~1B. Notwithstanding the large variance of $P(\sigma_j)$, only
supercritical networks (that is, with $\sigma > \sigma_c = 1$) have
self-sustained activity, $F>0$. As expected, critical networks have a
larger variance in the distribution of extinction times and present a
power law behaviour in the distribution of avalanche sizes with
exponent $3/2$ (not shown), in agreement with findings in biological
networks~\cite{Beggs03}.

Two types of oscillations are observed in this system. Under
sufficiently strong stimulation, {\em all\/} networks present
transient collective oscillations, with frequencies of the order of
the inverse refractory period (Fig.~1C). They are a simple consequence
of the excitable dynamics and the sudden synchronous activation by
stimulus initiation.  This transient behaviour is reminiscent of
oscillations widely observed in experiments~\cite{Laurent02}. Networks
with $\sigma> \sigma_{osc} > \sigma_c$ also present self-sustained
oscillations in the absence of stimulus (Fig.~1D), where
$\sigma_{osc}$ is a bifurcation threshold. The frequency depends on
the network parameters, but remain in the gamma range $30-60$~Hz. The
oscillations are similar to reentrant activity found in spatially
extended models of electrically coupled networks~\cite{Lewis01}, from
which analytical techniques for calculating the frequency could
perhaps be borrowed~\cite{lewis2000}. We remark that the large time
window used in the definition of $F$ is only chosen for convenience,
as it can be seen from Figs.~1C,D and 200~ms is sufficient to give
reliable averages. 

As a function of the stimulus intensity $r$, networks have a minimum
response $F_0$ (= 0 for the subcritical and critical cases) and a
maximum response $F_{max}$. We define the dynamic range
$\Delta=10\log(r_{0.9}/r_{0.1})$ as the stimulus interval (measured in
dB) where variations in $r$ can be robustly coded by variations in
$F$, discarding stimuli which are too weak to be distinguished from
$F_0$ or too close to saturation. The range $[r_{0.1},r_{0.9}]$ is
found from its corresponding response interval $[F_{0.1},F_{0.9}]$,
where $F_x = F_0 + x(F_{max}-F_0)$ (see Fig.~2C). This choice of a
10\%-90\% interval is arbitrary, but is standard in the literature and
does not affect our results.

As can be seen in Figs.~2A and 2B, the response curves $F(r)$ of the
networks present a strong enhancement of dynamic range compared with
the uncoupled case $\sigma= 0$. In the subcritical regime, sensitivity
is enlarged because weak stimuli are amplified due to activity
propagation among neighbours. As a result, the dynamic range
$\Delta(\sigma)$ increases monotonically with $\sigma$. In the
supercritical regime, the spontaneous activity $F_0$ masks the
presence of weak stimuli, therefore $\Delta(\sigma)$ decreases. The
optimal regime occurs precisely at the critical point (see
Fig.~2D). This is a new and important result, because it is perhaps
the first clear example of signal processing optimization at a phase
transition, making use of a standard and easily measurable performance
index.

\begin{figure}[!h]
\includegraphics[width=\columnwidth]{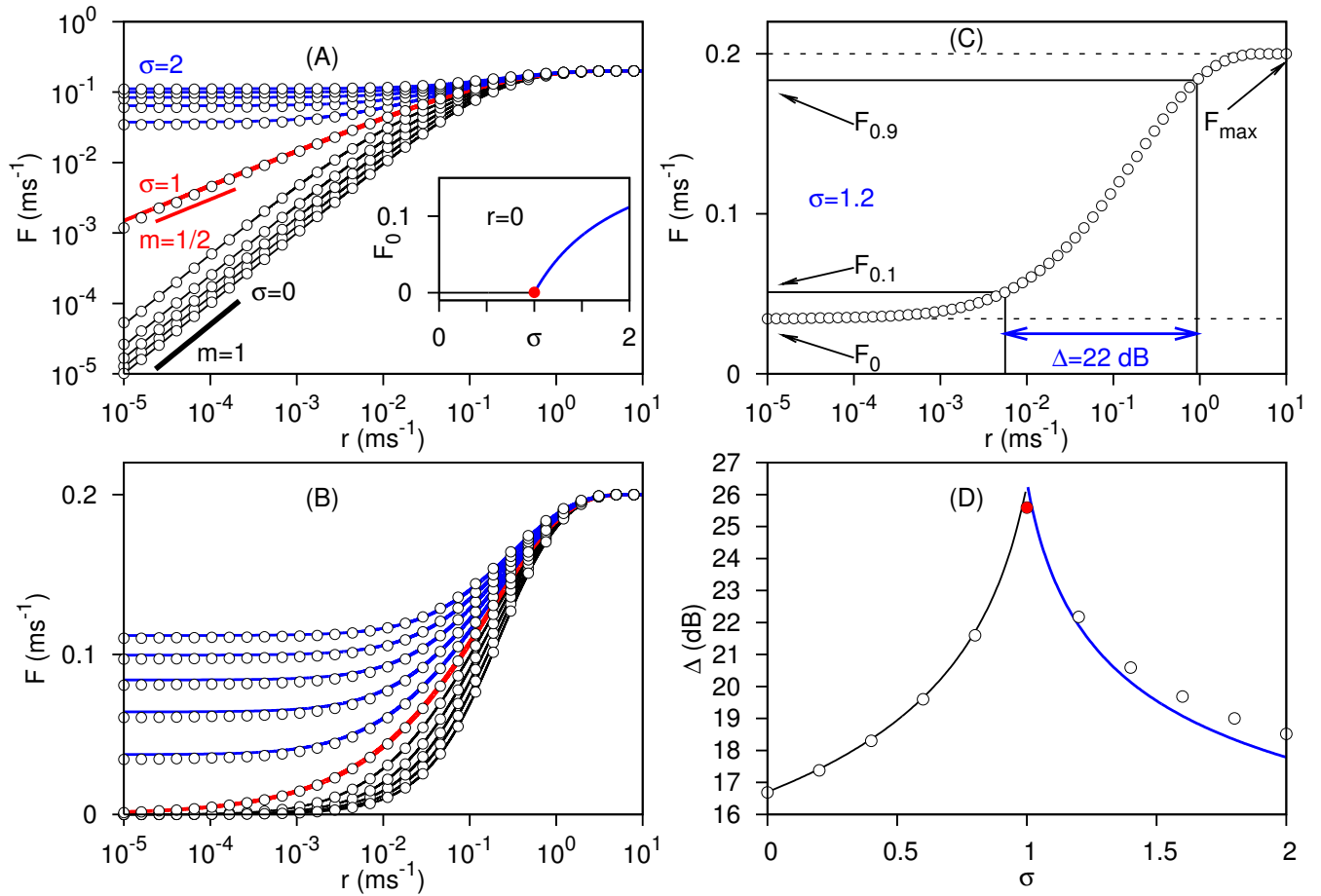}
\caption{Response curves and dynamic range. Points represent
simulation results with $N = 10^5$ sites, $K = 10$, $n = 5$ states and
$T = 10^3$~ms, whereas lines correspond to the mean-field model
described in the text. (A) Response curves (mean firing rate
vs. stimulus rate) from $\sigma = 0$ to $\sigma = 2$ (in intervals of
0.2). Line segments are power laws $F \propto r^m$ with $m = 1$
(subcritical) and $m = 1/2$ (critical).  Inset: spontaneous activity
$F_0$ vs. branching ratio $\sigma$. (B) The same as in (A), but with a
linear vertical scale. (C) Response curve for $\sigma = 1.2$ and
relevant parameters for calculating the dynamic range $\Delta$. (D)
Dynamic range vs.  branching ratio is optimized at the critical point
$\sigma = 1$.}
\end{figure}

The curves $F(r)$ could be fitted by a Hill function, but are not
exactly Hill. The theoretical curves in Fig.~2 are obtained from a
simple mean-field calculation (see below) that provides a very good
fit to the simulation data without free parameters, and correctly
predicts the exponents governing the low-stimulus response $F\propto
r^m$. An important point is that the Stevens-Hill exponent $m$ changes
from $m = 1$ in the subcritical regime to $m = 0.5$ at criticality. If
we assume that biological networks work in the optimal regime, the
critical value $m = 0.5$ suggests how exponents less than one could
emerge in psychophysics~\cite{Stevens} and neural
responses~\cite{Wachowiak01,Fried02}. We note that apparent exponents
between 0.5 and 1.0 are observed~\cite{Furtado06} if finite size
effects are present, that is, if $N$ is small.

In the simple mean-field approximation, we take $K_i = K$ and $p_{ij}$ as the
average value $\sigma/K$. The probability $p_t$ that an inactive site at time $t$
will be activated in the next time step by at least one of its $K$
neighbours (a fraction $F_t$ of which is currently active) is simply
$p_t = 1-(1-\sigma F_t/K)^K$. This leads to the following mean-field
map: 
\begin{equation}
\label{mapacampomedio}
F_{t+1} = P_t(0)\lambda + P_t(0)(1-\lambda)p_t\; ,
\end{equation}
where 

\begin{equation}
\label{stationary}
P_t(0) = 1-(n-1)F_t
\end{equation} 
is the approximate probability of finding a site in the resting state
and $\lambda(r)=1-\exp(-r\Delta t)$. The first term on the right-hand
side of equation~\ref{mapacampomedio} corresponds to activation due to
external input, and the second term corresponds to activation due to
neighbour propagation. In the stationary regime, the response function
$F(r)$ is given by the solution of
\begin{equation}
\label{campomedio}
F = \left(1-(n-1)F\right)\left[ 1 - (1-\sigma
  F/K)^K\left(1-\lambda(r)\right) \right] \; .
\end{equation}
Note that equation~\ref{stationary} is only exact in the stationary
state, but the resulting map is consistent because the result of its
iterations coincides with the solution of equation~\ref{campomedio}
(in particular, $F\leq 1/n$, so that $P(0)\geq 0$). 

As is usual in the statistical physics of phase
transitions~\cite{Marro99}, we analyse two limits: the critical
behavior without an external field and the effect of a vanishing field
at the critical point. In the absence of external stimulus ($\lambda =
0$), in the limit $\sigma\to\sigma_c = 1$ ($F\to 0$), the order
parameter behaviour is $F(\sigma)\simeq (\sigma-1)/C$ where
$C=(n-1)+(K-1)/2K$, which gives, from the definition $F(\sigma)
\propto (\sigma - \sigma_c)^\beta$, a critical exponent $\beta =
1$. On the other hand, at the critical point $\sigma = \sigma=1$ we
have $F(r) \approx \sqrt{r/C}$ which gives the Stevens exponent $m =
1/2$. Notice that the rate $r$ plays the role of an external field $h$
and, from the usual statistical-physics definition~\cite{Marro99}
$F\propto h^{1/\delta_h}$ we get the critical exponent $\delta_h=
2$. As expected, these are the classical mean-field exponents for
branching processes, which seem to be valid for our model even with
the presence of quenched disorder. An important conceptual point is
that the Stevens-Hill exponent is indeed found to be indeed a
statistical-physics critical exponent $m = 1/\delta_h$.


Optimization of dynamic range at criticality is very robust; the shape
of the plot in Fig.~2D does not depend on parameters such as the
average number of neighbours $K$ or refractory period $n$. The
random-network case seems to be a lower bound for the dynamic-range
enhancement. In future works, we will report that $\Delta(\sigma)$ is
even more enhanced in low-dimensional networks, and presents
non-rational Stevens exponents. Perhaps a better compromise between
larger dynamic range and biological realism would be a small-world or
scale-free network.

As an example of the robustness of the results, if we put
Hodgkin-Huxley elements with coupling with values $p_{ij}=0$ or 1,
which would correspond to a deterministic case with the presence of
absence of bonds, we obtain a $\Delta(p)$ curve, where $p$ is the
probability that a bond exists between the elements. This $\Delta(p)$
curve is very similar to the $\Delta(\sigma)$ curve of the present
model, that is, $\Delta(p)$ has a strong peak at the percolation phase
transition $p=p_c$ (to be reported elsewhere). However, the present
model has the virtue of enabling analytical results that provide a
benchmark for the performance of networks with other topologies.

Now we discuss the possible relevance of our results to biological
sensory processing.  Recent findings show that projection cells in
sensory systems are coupled via dendro-dendritic electrical synapses,
for example $\alpha$-ganglionar cells in the
retina~\cite{Schubert05,Hidaka05} and mitral cells in the olfactory
bulb~\cite{Kosaka05b,Migliore05,Christie05}. In most of these findings
the electrical coupling is mediated by connexins, but pannexins could
also be present, and could even be more important than connexins for
providing electrical coupling between excitatory cells~\cite{Vogt05}.
However, the functional role of this electrical coupling is largely
unknown. We propose that the electrically coupled dendritic trees in
these systems form an excitable network (where each element of our
model represents an excitable dendritic patch). Our results are
consistent with the reduction in sensitivity, dynamical range and
synchronization recently observed in retinal ganglion cell response of
connexin-36 knockout mice~\cite{Deans02}.

In the case of olfactory system, we identify the excitable random
network with the dendro-dendritic network in the
glomeruli~\cite{Kosaka05b}, and each element is interpreted as an
active dendritic compartment containing ion channels. It is known that
relevant electrical coupling between mitral cells is done at the
glomerular level~\cite{Migliore05,Christie05} because only cells that
have their apical dendrite tufts in the same glomerulus show
synchronized activity. In connexin-36 knockout mice, the synchronized
activity of mitral cells is absent. 

Our hypothesis could be tested in the following way. The dynamic range
of glomeruli is of the order of 30~dB as measured
recently~\cite{Wachowiak01,Fried02} (in contrast to 10~dB of single
olfactory receptor neurons). We predict that in connexin-36 knockout
mice, this dynamic range shall be strongly reduced. Of course, for a
decisive test, we need to examine if other electrical synapses based
on connexin-45~\cite{Zhang02} and pannexins~\cite{Vogt05} are
irrelevant in olfactory glomeruli.


	The text-book account of large dynamic range in intensity
coding is the recruitment model or its variants~\cite{Cleland99} where
different elements, with diverse activation thresholds, are
sequentially recruited. Our mechanism is not incompatible with this
scenario, and we expect that threshold variability in our model
enlarges the dynamic range even more. However, we must remember that
recruitment is a linear mechanism where, to explain each order of
magnitude in dynamic range, we must postulate a corresponding order of
magnitude in activation thresholds, which is not a plausible
assumption~\cite{Cleland99}.

	We may ask how the network could self-organize to the critical
point $\sigma_c=1$. It is not hard to conceive that homeostatic
mechanisms, acting on the number and conductance of gap junctions,
could tune the system. It is well known that extensive
pruning of gap junctions occurs during development and
maturation. This could represent the initial self-organization process
towards criticality. Spontaneous activity in the absence of input
furnishes a signature of supercriticality that could be used as a
feedback signal to control the system. 

This self-tuning criticality has recently been proposed in a model of
sound nonlinear amplification by Hopf oscillators in the
cochlea~\cite{Camalet00}. In the cochlear model, it is assumed that
each oscillator is poised at its Hopf bifurcation point, producing
enhanced sensitivity and enlarged dynamic range. The principal
difference from our mechanism is that it is a model for individual
cells, that is, it is not based on a collective phenomenon. This
implies that the dynamic-range exponent must be classical
(rational). In our case, the rational exponent $m=1/2$ is a
particularity of the random network, and, similarly to other
statistical-mechanics models, more-structured network topologies may
have non-classical exponents.

	Although in this work we restricted our attention to sensory
processing, we note that the dynamic range of more central networks
could also be improved by the same mechanism. Similarly to excitatory
networks, inhibitory networks must also work robustly in the presence
of large variations in input. So, the presence of electrical synapses
in cortical inhibitory networks~\cite{Sohl05} could reflect the same
principle. 

The mechanism for amplified nonlinear response due to wave creation
and annihilation is a basic property of excitable media. We found in
this work that, if active media are tuned at the critical point of
activity propagation, the response is optimized.  We proposed that
this principle is present in electrically coupled excitable dendritic
networks in projection neurons of sensory systems and is a generative
mechanism for psychophysical laws. This computational principle based
in critical activity could also be present in other brain regions and
could be implemented in artificial sensors by using excitable media as
detectors.

\begin{acknowledgments}
This research is supported by CNPq, FACEPE, CAPES and PRONEX. The
authors are grateful for discussions with A. C. Roque, R. F. Oliveira, D. Restrepo, T. Cleland, V. R. Vitorino de
Assis and for encouragement from N. Caticha.
\end{acknowledgments}


\end{document}